\documentclass[preprint,showpacs,nofootinbib,prd,aps]{revtex4-1}
\usepackage[utf8]{inputenc}
\usepackage{graphicx,amstext,amssymb}

\begin{document}
\title{Analysing direct photon spectra and elliptic flow from heavy ion collision measurements at the top RHIC energy within the integrated hydrokinetic model}

\author{V.~Yu. Naboka$^1$}
\author{Yu.~M.~Sinyukov$^{1}$}
\author{G.~M.~Zinovjev$^1$}
\affiliation{$^1$Bogolyubov Institute for Theoretical Physics,
03680 Kiev,  Ukraine}
%\\
%$^2$ExtreMe Matter Institute EMMI, GSI~Helmholtzzentrum f\"ur~Schwerionenforschung,
%D-64291 Darmstadt, Germany}

\begin{abstract}
The integrated HydroKinetic Model (iHKM) is applied to analyse the results of direct photon spectra and elliptic flow measurements in 200A GeV Au+Au collisions at RHIC for the three centrality bins. We detect the strong centrality dependence of photon elliptic flow as $v_2(p_T)$-coefficient increases towards peripheral collisions. The photon production in the model is accumulated from the different sources along with the process of relativistic heavy ion collision developing. Those include the primary hard photons from the parton collisions at  very early stage of the process, the photons generated at the pre-thermal phase of matter evolution, then thermal photons at equilibrated quark-gluon stage together with radiation displaying a confinement and, finally, from the hadron gas phase. Along the way a hadronic medium evolution
is treated in two distinct, in a sense opposite, approaches: chemically equilibrated and chemically
frozen system expansion.  We find that similar as it was found in iHKM for the LHC energies, a description of the direct photon spectra and elliptic flows is significantly improved if an additional portion of the photon
radiation, that is associated with hadronization processes, is included into consideration.

\end{abstract}

\maketitle

\section{Introduction}
Electromagnetic radiations is an unique messenger while probing new state of produced matter -- quark-gluon plasma (QGP) -- in relativistic nucleus-nucleus collisions \cite{Fein, Shur, Ru}. Nowadays it is recognized, photon spectra provide an information on the state of produced system just at the moment of photon
radiation and, hence, can even test some of the QCD calculations. At the same time it is necessary to take into account the significant influence on photon spectra that relativistic collective flows have in expanding superdense matter. Typically, such flows are described by hydrodynamic-based models containing different stages of the matter evolution in relativistic A+A collisions. Extensive theoretical studies of the photon production in such models inspired by the unexpectedly large direct photon yield as well as their elliptic flow  measured by PHENIX Collaboration at RHIC in the recent years \cite{RHIC_photon} have considerably
increased the possible number of photon radiation sources in order to resolve "direct photon flow  puzzle" \cite{Renk, Sriva, Paquet}.

    In our recent paper \cite{ihkm-photon} we analyzed within the integrated HydroKinetic Model (iHKM) \cite{iHKM} the ``photon puzzle'' observed by the ALICE Collaboration at CERN \cite{ALICE_spectra}. The model describes well the bulk hadron observables at different centralities: various particle yields and number ratios, pion, kaon, (anti)proton spectra, elliptic flows, pion and kaon interferometry radii at the top RHIC energy 200A GeV  Au+Au collisions and both LHC  energies for Pb+Pb collisions: 2.76 TeV and 5.02 TeV per nucleon pair \cite{ihkm-rhic,ihkm-lhc276, ihkm-lhc502}. The  predictions of the model, e.g.  \cite{kaon-femto, Kstar}, were later confirmed in the experiments, e.g. \cite{ALICE-kaon}. The iHKM parameters  adjusted for baryon observables are kept the same in all simulations,  except for the initial maximal energy density and baryon chemical potentials which depend on the collision energy at RHIC and LHC.  Using the corresponding parameters, in Ref. \cite{ihkm-photon} the iHKM has been applied to describe the direct photon transverse spectra and their anisotropy, expressed through the so-called $v_2$-coefficients, at the LHC energy $\sqrt{s_{NN}}= 2.76$ TeV \cite{ALICE-kaon}.
		
		The direct photon production accumulates the process of quark collisions at the very early stage resulting in the so-called prompt photons, photons produced at the pre-thermal stage, thermal photons from QGP and  photons from the hadronic stage of the matter evolution. It was found that the data description, both the photon spectra and $v_2$, can be improved significantly if one takes into account an additional (hypothetical) photon radiation originated from the hadronization process.
		
		The nature of the photon radiation during hadronization is not clear yet, in literature  a few possible mechanisms are presented \cite{azimuth-aniz, Kharz, Camp-1,
Pratt, Itakura}, e.g. such as "magnetic
bremsstrahlung-like radiation" (synchrotron radiation) deconfinement phase \cite{azimuth-aniz}  and  electromagnetic radiation that accompany formation of hadrons from quarks \cite{Pratt}.  We use such an idea supposing that a
specific photon radiation takes place during the process of hadronization. Trying to give more credibility to such mechanisms we speculate here adapting the
phenomenological  prescription for describing the photon emission from the hadronization space-time layer in the cross-over scenario at the top LHC and LHC energies. The  model of such additional photon radiation has been developed
in Ref. \cite{ihkm-photon}. Since a matter of a space-time layer where a hadronization takes place is actively involved in anisotropic transverse flow, both positive contributions to the spectra and $v_2$ are considerable albeit such a possible phenomena needs a further research and elaboration.
   In this paper we apply this model also for the top RHIC energy.
	
The paper is organized as follows. Section \ref{sec:iHKM} is devoted to the brief review of iHKM in its application for modeling the matter
evolution.  The description of  different sources of the direct photon radiation associated with the corresponding stages of the matter evolution is given in Section
\ref{sec:sources}. The results and discussion are presented in Section \ref{sec:results}. The Summary is given in Section \ref{sec:summary}.

\section{Integrated hydrokinetic model}
\label{sec:iHKM}
Before addressing the possible sources of photon emission, we will describe briefly the model of matter evolution, used in this research. We utilize the integrated hydrokinetic model (iHKM) \cite{iHKM} for modelling evolution of heavy ion collision. This model was developed and used for higher energies (mostly, for energies available at Large hadron Collider), but it is also applicable for RHIC energies. iHKM considers heavy ion collision as consisting of five stages: initial stage generation, prethermal stage, thermal (hydrodynamic) stage, particlization stage and UrQMD stage. Further we describe each stage more specifically.

\subsection{Initial state generation}

For initial state generation we use GLISSANDO \cite{gliss,gliss2} package, which works in the frame of semiclassical Glauber model. The output initial state of GLISSANDO package is then attributed to initial time $\tau_0=0.1$ fm/c  \cite{iHKM,ihkm-rhic}. Initial state is set by the combination
of several factors. These factors are: 1) coefficient $\alpha$, which defines a relative contribution of wounded nucleons and binary collisions to formation of initial energy density profile, 2) $\Lambda=100$ sets the momentum spectrum anisotropy of partons at the initial moment of time according to the color glass condensate (CGC) model \cite{iHKM,ihkm-rhic}. In this paper we use $\alpha=0.18$ as a coefficient describing the dependence of multiplicity on collision centrality for the top RHIC energy for Au+Au collisions {ihkm-rhic}.

\subsection{Relaxation model of prethermal stage}

Relaxation model is the model of initial state thermalization \cite{preth,relaxation, iHKM}. It models the continuous transition from locally non-equilibrated initial state in moment $\tau_0$ (which is equal to $0.1$ fm/c in our calculations) to the near equilibrated state in the moment $\tau_{th}$ (the same as for LHC case, this model parameter is put equal to $1.0$ fm/c). Further evolution of equilibrated matter can be described by (viscous) relativistic hydrodynamics. The energy-momentum tensor of matter on the prethermal stage can be written as
\begin{eqnarray}
T^{\mu \nu}(x)=T^{\mu \nu}_{\text{free}}(x){\cal
P}(\tau)+T_{\text{hydro}}^{\mu \nu}(x)[1-{\cal P}(\tau)], \label{tensor}
\end{eqnarray}
where $T^{\mu \nu}_{\text{free}}(x)$ is the free-streaming part of the total energy-momentum tensor and $T^{\mu \nu}_{\text{hydro}}(x)$ is hydrodynamically evolving component. The function $\cal P(\tau)$ has the form \cite{preth,relaxation}
\begin{eqnarray}
{\cal P}(\tau)=  \left(
\frac{\tau_{\text{th}}-\tau}{\tau_{\text{th}}-\tau_0}\right
)^{\frac{\tau_{\text{th}}-\tau_0}
 {\tau_{\text{rel}} }}.
 \label{P}
\end{eqnarray}
As one can see, $P(\tau_0)=1$ and $P(\tau_{th})=0$. Thus, the matter is gradually transiting from the pure free-streaming in the initial time $\tau_0$ to pure hydrodynamic evolution at the thermalization time $\tau_{th}$. Parameter $\tau_{rel}$ describes the rate of this transition. As for the LHC energy, for RHIC we put $\tau_{rel}=0.25$ fm/c as in Refs. \cite{iHKM, ihkm-rhic}. Writing down the conservation laws for the total energy-momentum tensor and accounting for the properties of the free-streaming energy-momentum tensor, we have
\begin{eqnarray}
\partial_{;\mu}\widetilde{T}^{\mu
\nu}_{\text{hydro}}(x)= - T^{\mu \nu}_{\text{free}}(x)\partial_{;\mu}
{\cal P}(\tau), \label{pre-equation}
\end{eqnarray}
where $\widetilde{T}^{\mu\nu}_{\text{hydro}}(x)=[1-{\cal P}(\tau)]T^{\mu\nu}_{\text{hydro}}(x)$. The relaxation model in the form used in this paper calculations, also contains a shear viscosity tensor terms in the Israel-Stewart form. Evolution of the shear stress tensor can be derived analogically to (\ref{pre-equation}):
\begin{eqnarray}
[1-{\cal P}(\tau)]\left \langle u^\gamma \partial_{;\gamma}
\frac{\widetilde{\pi}^{\mu\nu}}{(1-{\cal P}(\tau))}\right \rangle
=-\frac{\widetilde{\pi}^{\mu\nu}-[1-{\cal
P}(\tau)]\pi_\text{NS}^{\mu\nu}}{\tau_\pi}-\frac {4}{3}
\widetilde{\pi}^{\mu\nu}\partial_{;\gamma}u^\gamma.
\label{pre-viscous}
\end{eqnarray}

If $\tau \rightarrow \tau_{th}$ then ${\cal P} \rightarrow 0$ and system reaches the  target energy-momentum tensor in the  Israel-Stewart form.

\subsection{Hydrodynamic stage}

The stage of locally equilibrated hydrodynamic evolution follows the prethermal stage at $\tau=\tau_{th}$ and lasts till the hypersurface of constant temperature $T=165$ MeV. By setting the source terms in (\ref{pre-equation}) and (\ref{pre-viscous}) to zero, we get the casual viscous hydrodynamic evolution equations. For viscosity coefficient on this and previous stage we use its minimal value, $\eta/S=0.08$. The Laine-Schroder equation of state (EoS) \cite{Laine} is used on this stage (and on prethermal stage as well). The reason to choose this EoS is the possibility of continuous transition to hadron gas. The difference from the LHC case is the necessity to  use for top RHIC energy the different chemical potentials into consideration. We settle chemical potentials equal to be the same as in Ref. \cite{ihkm-rhic}  and use the same prescription to account them as in that paper.
It is worthy note that the change EoS to another one, e.g. EoS from Hot QCD Collaboration \cite{EoS2} does not destroy the results, if simultaneously one provides an appropriate adjusting of the initial time for the superdense matter formation and related maximal initial energy density \cite{lhc502}.

\subsection{Particlization}

As far as hydrodynamic model is not applicable when matter becomes diluted, we have to make a transition from hydrodynamic stage to further evolution of matter as hadron-resonance gas. We make this transition at a hypersurface of constant temperature, specifically, $T=165$ MeV at the Laine-Schroder EoS (it corresponds to energy density $\epsilon = 0.5$ GeV/fm$^3$). As noted before, using the Laine-Schroder EoS ensures the continuous transition on this stage. For building the switching hypersurface we utilize the Cornelius routine \cite{convert}. For conversion of fluid matter to hadron gas, we use the Cooper-Frye formula, also Grad's 14-momentum ansatz is used to apply viscous corrections.

\subsection{Hadron gas stage}

The further evolution of matter as hadron gas is simulated by Ultrarelativistic Quantum Molecular Dynamics (UrQMD) \cite{urqmd} in the original iHKM. However, this approach proves to be insufficient for the case of photon spectra research, as UrQMD is not appropriate for describing the photon emission. To solve this problem, we use the two different approaches, similar as it is done in  Ref. \cite{ihkm-photon}. First one is rather simple -- we just prolongate the chemically equilibrated evolution also in the hydrodynamic stage to the lower temperature, such as $T=100$ MeV. While this approach seems to be rather primitive, it can provide an approximate estimate of photon emission and anisotropy. The second approach, on the other hand, is to use the original hydrokinetic model (HKM) \cite{original HKM, HKM}, which describes a continuous particle generation after hadronization temperature $T=165 MeV$, and provides the opposite approximation -- a chemically frozen matter evolution.

\section{Sources of photon emission}
\label{sec:sources}
There are many different sources of photons emitted during the heavy ion collision, each of them having the most impact during a certain stage of the collision process. Overall, they can be divided into prompt photons, which are emitted in the earliest stage of collision before thermalization; thermal photons, emitted from the thermalized quark-gluon plasma and hadron gas; and finally, photons emitted
at the hadronization stage, which were discussed in our previous paper for LHC \cite{ihkm-photon} We will discuss these photon sources in more detail.

\subsection{Prompt photons}

 Prompt photons are photons resulting from the very first instances of the nuclear collision, mainly through the Compton scatterings of partons and quark-antiquark annihilations. These photons can be calculated using the perturbative QCD (pQCD) techniques. Accounting the results of various experiments \cite{ATLAS_photon, CMS_photon}, it could be stated that prompt photon spectra scale with the binary nucleon-nucleon collision number. Thus, prompt photon spectra in Au+Au collisions can be presented as convolution of binary collision number $N_{coll}$ with the proton-proton spectra, calculated using pQCD. The latter can be presented in the form of cross section

\begin{equation}
d\sigma = \sum_{i,j,k}f_i\otimes f_j \otimes d\hat{\sigma}(ij\rightarrow k)\otimes D^{\gamma}_k,
\label{g_s}
\end{equation}
where the summation runs over all possible partonic subprocesses, $f_i$ and $f_j$ are the parton
distribution functions, $D^{\gamma}_k$ is the fragmentation function and $d\hat{\sigma}$ ~---
cross section of the corresponding partonic subprocess. We calculate the binary collision number \cite{Ncoll} with the Monte Carlo Glauber code \cite{gliss, gliss2}. The cross section in eq. (\ref{g_s}) is calculated by perturbative expansion in the strong coupling constant. The QCD scales, which are present in such an expansion, are set to $Q_{fact}=Q_{ren}=Q_{frag}=0.5~p_T$. In order to compute the cross section (\ref{g_s}) for momentum range $0.5$ GeV/c $< p_T < 6$ GeV/c, we address the JETPHOX package \cite{JETPHOX}. The pQCD calculations have rather high lower $p_T$-limit, especially for such low proportionality coefficient in coupling constant/momentum dependence as $Q=0.5~p_T$, which we have chosen in our calculations. Thus, we address to the fact that we can make our calculations for higher proportionality coefficient, specifically $Q=8.0~p_T$, and then just rescale the obtained spectra (see more information on this method in Ref. \cite{Paquet}). In our calculations we utilize EPS09 parton distribution functions \cite{EPS09} and BFG II fragmentation functions \cite{BFG-2}, both of which are represented as tables of values.

\subsection{Prethermal and thermal photons}

As far as iHKM contains prethermal stage, it is natural that this stage also provides some photon emission, which should be treated differently from the photons originating from thermalized QGP. These photons also should not be mistaken for prompt photons, which are generated in the first instances of collision. Prethermal photons in our approach are emitted by the matter at times $0.1$ fm/c $< \tau < 1.0$ fm/c. For the prethermal stage of iHKM, total energy-momentum tensor consists of hydrodynamic and free-streaming components (\ref{tensor}), so we consider prethermal photons as coming from the hydrodynamic component. The resulting spectra is then multiplied by $1-{\cal P(\tau)}$, as this is a weight coefficient of hydrodynamic component in the total energy-momentum tensor.

Spectra of photons originated from QGP was described long ago \cite{Arnold} in the leading order of strong coupling constant $g_s$. These formulas are used in this paper calculations. The main sources of such photon emission are leading order 2 $\rightarrow$ 2 processes in the hot QGP. It also should be noted that next to leading order processes can make a contribution of the same order to the spectra \cite{Gelis, Ghiglieri}. For describing the spectra and anisotropic flow of QGP-originating thermal photons one needs a reliable tool for modelling the matter evolution. As such, we utilize the above mentioned iHKM (\ref{sec:iHKM}) The photon emission in the relativistic-invariant form can be described by the formula

\begin{equation}
k^0\frac{d^7N}{d^3\textbf{k}d^4\textbf{x}}={k}\cdot {u}~\frac{\nu_e(k\cdot {u})}{(2\pi)^3},
\label{spectra2}
\end{equation}

where $\textbf{k}$ is the photon momentum and $\textbf{u}$ is the collective flow of matter in the coordinate $\textbf{x}$. Thus, $\nu_e(k\cdot {u})$ has a physical meaning of spontaneous emission rate of photons with momentum $\textbf{k}$. The evaluation of $\nu_e(k\cdot {u})$ was made in \cite{Arnold} by summarizing emission output from partonic 2 $\rightarrow$ 2 subprocesses with inclusion of near-collinear bremsstrahlung and inelastic pair annihilation contributions, and accounting for Landau-Pomeranchuk-Migdal suppression effects. The final spectra then has the form
\begin{equation}
\frac{dN}{2\pi k_Tdk_T}=\sum_i\frac{1}{(2\pi)^4} \Delta^4 V(x_i)\int_0^{2\pi} {k}\cdot {u(x_i})\nu_e({k}\cdot {u(x_i}))d\phi,
\label{spectra_qgp}
\end{equation}
$\Delta^4 V=\tau \Delta \tau\Delta x\Delta y\Delta\eta$ is a volume of a cell in the computation grid. The summation runs over a 3-dimensional computation grid. As far as this formula is applicable for QGP, it runs over all cells with $T>165 MeV$, thus affecting prethermal and hydrodynamic stages of iHKM (\ref{sec:iHKM}).

The emission from cells with lower temperature,  $T<165 MeV$ is described by the same formula (\ref{spectra_qgp}), but with $\nu_e(k\cdot {u})$ derived for hadron gas emission. The hadron gas photon emission rate consists of many terms, which have a different nature. These include: 1) meson gas photon emission \cite{Turbide}; 2) emission with a specific behaviour of the $\rho$-meson self-energy \cite{Heffernan}; 3) emission originating from reactions $\pi+\rho\rightarrow\omega+\gamma, \rho+\omega\rightarrow\pi+\gamma, \pi+\omega\rightarrow\rho+\gamma$ \cite{Rapp}; 4) photon emission from $\pi\pi$ bremmstrahlung \cite{Heffernan}. As it was pointed above, we use two different, in some sence - opposite, scenarios: chemically equilibrated and chemically frozen evolution of hadronic matter to see a difference in the corresponding impact to the final results.

\subsection{Hadronization stage photons}

In our previous work \cite{ihkm-photon} for LHC energies we have developed a prescription  for calculating the photon emission on the hadronization stage has been developed. This phenomenological approach is based on a suggestion about additional photon radiation from a confining process proposed in Refs.  \cite{azimuth-aniz,Kharz, Camp-1,
Pratt, Itakura}. Not touching the details of various mechanisms that  were considered in these papers, we use a following phenomenological prescription \cite{ihkm-photon}.

Let $G_{hadr}(t,{\bf r},p)$ be the emission function of the additional photon radiation at the hadronization stage:
\begin{equation}
\frac{d^3N_{\gamma}}{d^3p}=\int dtd^3r~G_{hadr}(t,{\bf r},p)
\label{G_hadr}
\end{equation}
Let $\sigma$ be the hypersurface of temporal points $t_{\sigma}({\bf r}, p)$ of maximal emission for photons with momentum $p$. Let us change variables from $({\bf r}, p)$ to $({\bf x}, p)$, where $({\bf x}$ includes the saddle point for given $({\bf r}$ \cite{var-1,var-2}: ${\bf x}={\bf r}+\frac{{\bf p}}{p^0}t_{\sigma}({\bf r},p)$. Then use the saddle point approximation for emission function $G_{hadr}\approx F(t,{\bf x},p)\exp(-\frac{(t-t_{\sigma})^2}{2D^2})$, where $F$ has smooth dependence on $t$. Then
\begin{equation}
\frac{d^3N_{\gamma}}{d^3p}=\int d^3x \left|1-\frac{{\bf p }}{p^0}\frac{\partial t_{\sigma}({\bf x}, p)}{\partial {\bf x}}\right|\int dt F(t_{\sigma}({\bf x}, p), {\bf x}, p) \exp\left(-\frac{(t-t_{\sigma}({\bf x}, p))^2}{2D_c^2(t_{\sigma}({\bf x}, p), {\bf x}, p)}\right)
\label{add}
\end{equation}
Assuming that the hypersurface of maximum photon emission corresponds to the hadronization isotherm ($T_h=165$ MeV in our model), we can write down (\ref{add}) in the invariant form:
\begin{equation}
p^0\frac{d^3N_{\gamma}}{d^3p}=\int_{\sigma_h}d^3\sigma_{\mu}(x)p^{\mu}F\left(p\cdot u(x),T_h\right)D_c\left(p\cdot u(x),T_h\right)\theta(d\sigma_{\mu}(x)p^{\mu}),
\label{cooper_frye}
\end{equation}
where $\theta(z)$ is the Heaviside step function, which excludes negative contribution to spectra from non-space-like parts of hadronization hypersurface. We try to include this synchrotron radiation mechanism in the simplest phenomenological form, so we suppose that $FD$ function in (\ref{cooper_frye}) has thermal-like form:
\begin{equation}
F D_c = d_c\gamma_{hadr}~f_{\gamma}^{eq}\left(p\cdot u(x),T_h\right)=\gamma_{hadr}d_c\frac{1}{(2\pi)^3}\frac{g}{\exp\left(p\cdot u(x)/T_h\right)-1},
\label{surface}
\end{equation}
where $p\cdot u \equiv p^{\mu}u_{\mu}$, $g=2$, $T_h = 165$. Value $\gamma_{hadr}$ is defined by the hadronization process, and $d_c \propto \left\langle D\right\rangle$ is defined by temporal width of this process.
We use the value $\alpha\equiv d_c\gamma_{hadr}=0.04$ as providing the best description for the $0-20\%$ centrality events.

\begin{figure}
     \centering
     \includegraphics[width=1.0\textwidth]{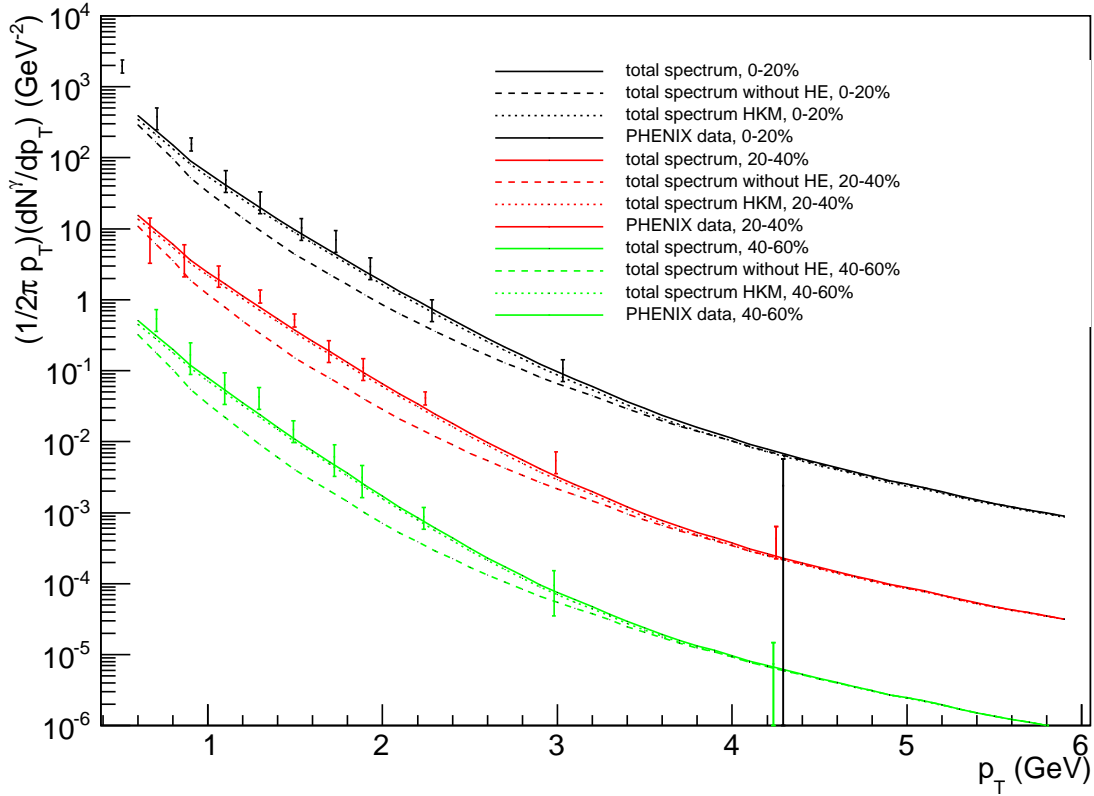}
     \caption{\small
Total photon spectra for the different models: iHKM chemically equilibrated with contribution from hadron emission (HE), iHKM chemically equilibrated without HE contribution, HKM chemically frozen at the hadron stage with continuous transition from hydrodynamics to hadron gas (with HE contribution). Results for centralities 0-20\%, 20-40\% and 40-60\% are included. Spectra for 0-20\% centrality are multiplied by factor of 100, and spectra for 20-40\% centrality are multiplied by factor of 10. Experimental results are taken from \cite{RHIC-spectra}.}
\label{fig:spectra_all}
\end{figure}

\begin{figure}
     \centering
     \includegraphics[width=1.0\textwidth]{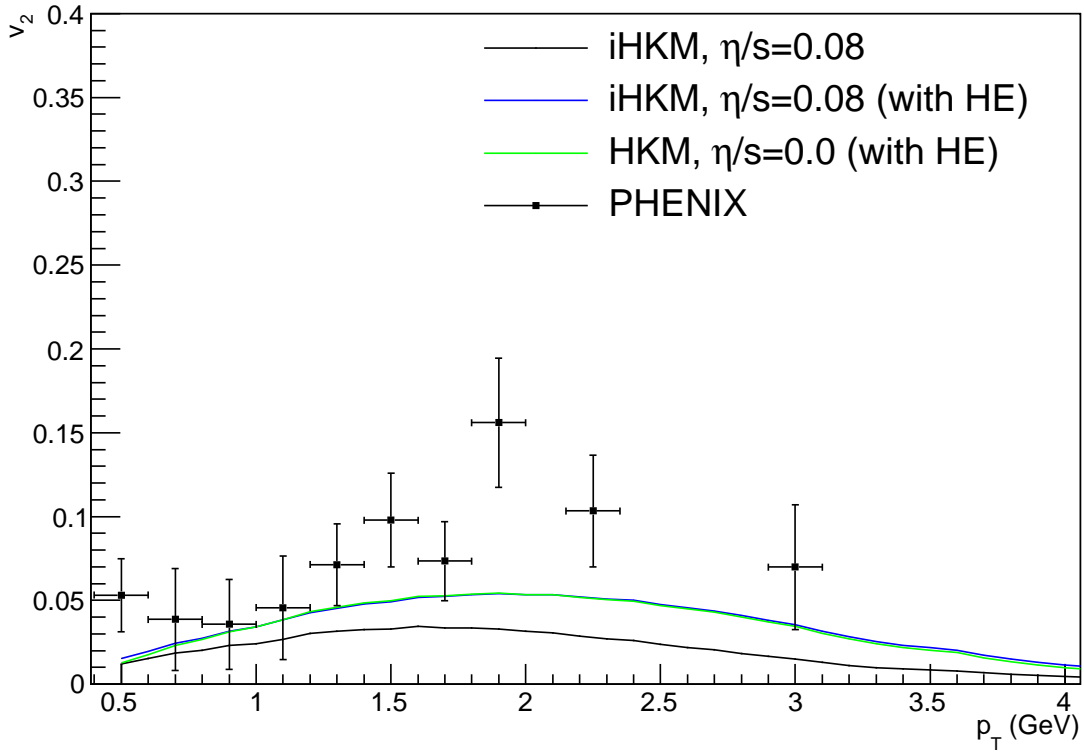}
     \caption{\small
Anisotropic flow for 0-20\% centrality for the different models: iHKM chemically equilibrated with  contribution, iHKM chemically equilibrated without hadronization emission (HE) contribution, HKM chemically frozen at the hadron stage with continuous transition from hydrodynamics to hadron gas (with HE contribution). Experimental results are taken from \cite{RHIC-v2}.}
\label{fig:v2-0-20}
\end{figure}

\begin{figure}
     \centering
     \includegraphics[width=1.0\textwidth]{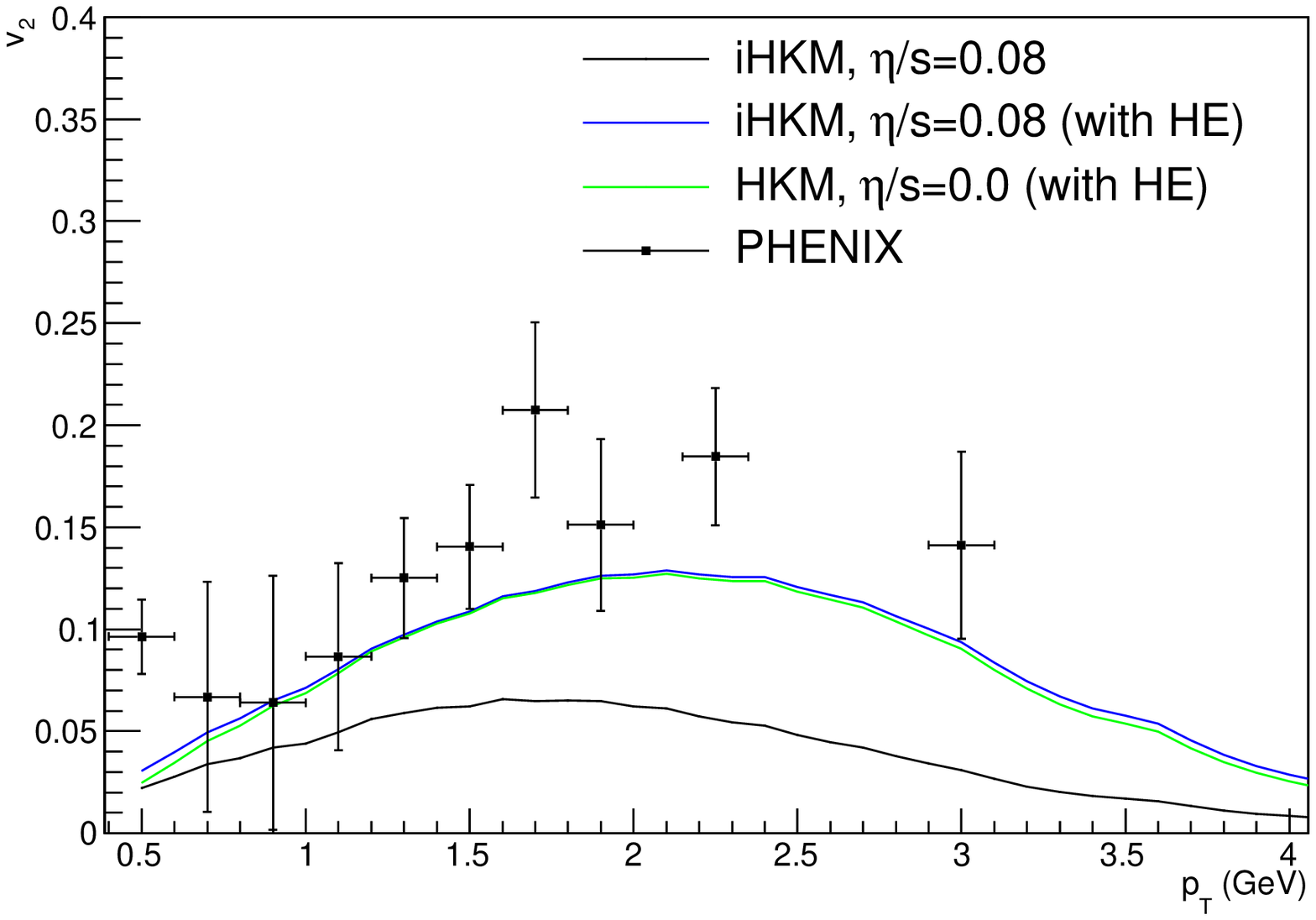}
     \caption{\small
Anisotropic flow for 20-40\% centrality for the same conditions as in Fig. \ref{fig:v2-0-20}}
\label{fig:v2-20-40}
\end{figure}

\begin{figure}
     \centering
     \includegraphics[width=1.0\textwidth]{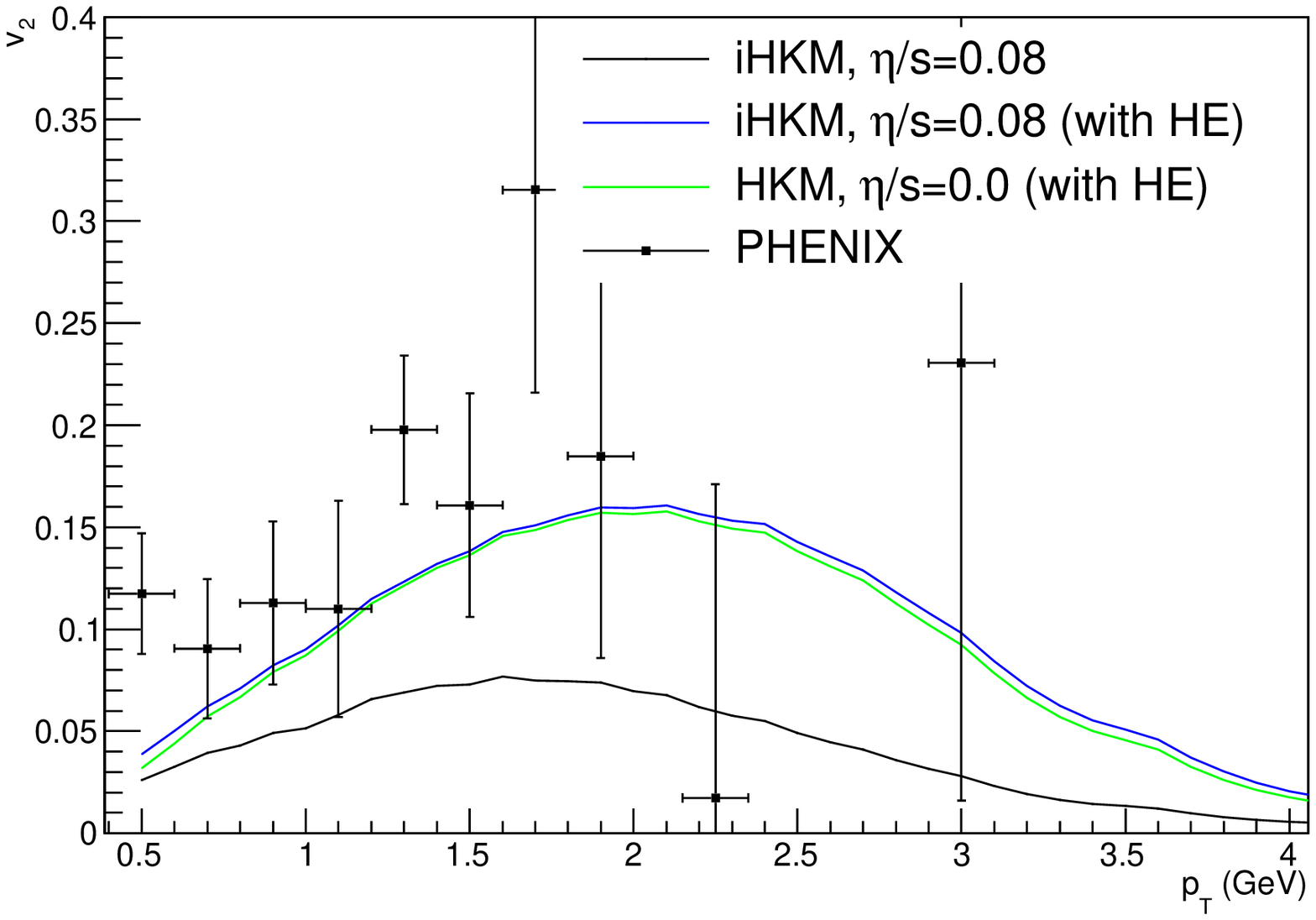}
     \caption{\small
Anisotropic flow for 40-60\% centrality for the same conditions as in Fig. \ref{fig:v2-0-20}}
\label{fig:v2-40-60}
\end{figure}

\section{Results and discussion}
\label{sec:results}

The total photon spectra is calculated as a sum of above mentioned constituents, specifically, prompt photon spectra, thermal photon spectra (which includes prethermal photon spectra as well because iHKM contains prethermal stage), and hadronization surface photons. We compare two main scenarios of evolution: chemically equilibrated and chemically frozen. For the former scenario, hydrodynamic stage lasts not only to hadronization temperature $T=165$ MeV, but is prolonged to as far as $T=100 MeV$ in order to provide necessary data to describe hadronic stage emission (which is included in thermal photon spectra). And chemically frozen scenario is described by the original hydrokinetic model (HKM) \cite{original HKM, HKM}. In this model below the temperature $T=165$ MeV, instead particles leave the hadron fluid continuously, matter evolution is neither thermal nor chemically equilibrated. In opposite to the first chemically equilibrium scenario, the evolution at this stage is chemically frozen ~--- of all the inelastic reactions, only the decays of resonances are allowed. Thus this approximation gives more realistic description of matter evolution between temperatures $T=100 MeV$ and $T=165 MeV$.

As for iHKM parameters, we use mostly the same set of parameters, which we have used for LHC \cite{iHKM} ($\tau_{0}=0.1$, $\tau_{th}= 1$ fm/c, $\tau_{rel}=0.25$ fm/c, $\Lambda=100$, $\eta/S=0.08$), however GLISSANDO initial state weight  coefficient is now $\alpha=0.18$ instead of 0.24 at the LHC energies $\sqrt{s_{NN}}=2.76$ and 5.02 TeV for $Pb+Pb$ collisions. It is better describers charged particle multiplicity/centrality dependence \cite{ihkm-rhic}. Of course, also the baryon chemical potential in mid-rapidity has now non-zero value \cite{ihkm-rhic}. Overall, this set of parameters is chosen because it provide a good description of bulk observables for different centralities, such as particle yields and number ratios, charged particle multiplicity, pion/kaon/(anti)proton spectra, charged particle momentum anisotropy and pion and kaon HBT-radii at RHIC \cite{ihkm-rhic}. Thus, one of the strongest point of our approach is using the same model and parameters to describe not only photon emission, but a set of bulk hadron observables as well.

The results are represented for three centrality classes: $0-20\%$, $20-40\%$ and $40-60\%$. The total photon spectra for all centralities are shown in Fig. \ref{fig:spectra_all}. As mentioned above, we compare two scenarios: chemically equilibrated (iHKM) and chemically frozen (HKM). For the first scenario, results without hypothetical hadronization emission are also demonstrated. The resulting spectra show that there is insignificant difference between two scenarios, while the surface emission gives large impact on the final results. This should not be a surprise, because, in comparison with analogical description for LHC \cite{ihkm-photon}, we have chosen a larger weight coefficient of surface emission $\alpha=0.04$. Such a choice was made because choosing such a large $\alpha$ does not lead to a significant worsening of total spectra for larger centralities, as opposed to LHC case. The photon anisotropic flow dependence on photon momentum is shown on Figs. \ref{fig:v2-0-20} - \ref{fig:v2-40-60}. Again, the two scenarios give a very close description, and surface emission gives a large impact on results in both scenarios.

\section{Summary}
\label{sec:summary}
In this paper we investigate the photon spectrum and its momentum anisotropy for heavy ion collisions at RHIC energy $\sqrt{s_{NN}}=200A$ GeV in the framework of the integrated hydrokinetic model. Incindentally, we treat different sources of photon emission in this investigation, precisely, prompt photons, photons radiating at the pre-thermal phase along its evolution, thermal photons from quark gluon plasma (including synchrotron radiation yield) and hadronic gas stages emitting because of hadronization process. The set of iHKM parameters used here is just the same as that we have handled for successful description of hadronic bulk observables in $Au+Au$ $\sqrt{s_{NN}}=200$ GeV collisions at RHIC \cite{ihkm-rhic} including the particle yields and number ratios, pion, kaon, proton, antiproton spectra, transverse momentum anisotropy, and HBT-radii.

We compare results of the two approaches to describe  the hadron gas emission supposing chemically equilibrated and chemically frozen evolution at the corresponding stage. Both approaches lead to quite similar results and they are mostly within the error bars for all centralities. However, it is shown that the photon emission mechanism allowing to probe a confinement plays an important role in correct description both spectra and anisotropic flow.

\section{Acknowledgments}
The research was carried out within the scope of the European Research Network EUREA: ``Heavy ions at ultrarelativistic energies'' and corresponding Agreement with  National Academy of Sciences (NAS) of Ukraine. It is partially supported by NAS  of Ukraine Targeted research program ``Fundamental research on high-energy physics and nuclear physics (international cooperation)'', Agreement F7-2018.

\end{document}